\documentclass[letterpaper]{article}
\usepackage{iccc}

\usepackage{times}
\usepackage{helvet}
\usepackage{courier}
\usepackage{graphicx}
\usepackage{multirow}
\usepackage{url}
\title{Deep Learning in a Computational Model for Conceptual Shifts in a Co-Creative Design System}
\author{Pegah Karimi$^1$, Mary Lou Maher$^1$, Nicholas Davis$^1$, Kazjon Grace$^2$\\
$^1$UNC Charlotte, $^2$The University of Sydney\\
$^1$USA, $^2$Australia\\
pkarimi@uncc.edu, m.maher@uncc.edu, ndavis64@uncc.edu, kazjon.grace@sydney.edu.au \\
}
\begin{document} 
\maketitle
\begin{abstract}
\begin{quote}
This paper presents a computational model for conceptual shifts, based on a novelty metric applied to a vector representation generated through deep learning. This model is integrated into a co-creative design system, which enables a partnership between an AI agent and a human designer interacting through a sketching canvas. The AI agent responds to the human designer's sketch with a new sketch that is a conceptual shift: intentionally varying the visual and conceptual similarity with increasingly more novelty. The paper presents the results of a user study showing that increasing novelty in the AI contribution is associated with higher creative outcomes, whereas low novelty leads to less creative outcomes.
\end{quote}
\end{abstract}

\section{Introduction}

Creative systems are computational systems that either model human creativity in some manner or are designed to support and inspire creativity. Over the last few years, three main approaches to these systems have emerged: fully autonomous creative systems, creativity support tools, and co-creative systems. Fully autonomous creative systems, part of the field of computational creativity, are designed to generate creative artifacts or exhibit creative behaviors \cite{colton2015painting,das2014poetic}. Creativity support tools, on the other hand, are technologies that can support human creativity by accelerating or augmenting some facets of the creative process \cite{shneiderman2007creativity,voigt2012towards}. Finally, co-creative systems incorporate concepts from both fully autonomous systems and creativity support tools: they enable human users and computer systems to work together on a shared creative task \cite{davis2015enactive,yannakakis2014mixed}.  

In this paper, we introduce the algorithms for a co-creative sketching tool called the Creative Sketching Partner (CSP), which involves collaboration between a designer and an AI agent on a shared design task. Figure 1 illustrates the CSP tool, in which the design task is described at the top and the three sketches below represent the responses to this task. The two sketches at the top represent the user's initial sketch on the left and the AI agent's responding sketch and label for the sketch on the right. The sketch at the bottom of the canvas is the user's new sketch, with the shaded region showing the user's additions inspired by the AI agent's sketch. The system utilizes a computational model of conceptual shifts \cite{karimi2018creative,karimi2018deep} to guide users toward different aspects of the design space based on the amount of visual and conceptual similarity to the user's sketch input. Visual similarity entails identifying a sketch that shares some structural characteristics, whereas conceptual similarity identifies a concept that has some semantic relationship. We present users with stimuli that have either both high visual and conceptual similarity (like a pen and a pencil) or low visual and conceptual similarity (like a dolphin and a chair).

\begin{figure}
    \centering
    \includegraphics[width=6cm]{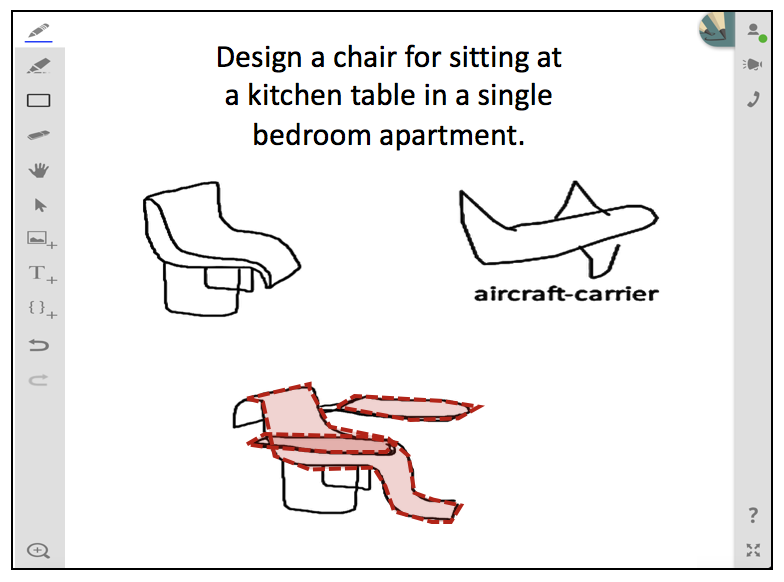}
    \caption{The Creative Sketching Partner interface.}
    \label{fig:my_label}
\end{figure}

\citeauthor{karimi2018evaluating} \shortcite{karimi2018evaluating} introduced a framework of ways to evaluate creativity in co-creative systems. It was found that current co-creative systems research tends to focus on measuring the usability of the system, rather than on operationalising creativity. This demonstrates an opportunity for adopting  metrics from computational creative systems in order to empower co-creative systems with the capacity to measure the creativity of their contributions to the output. For our conceptual shift model, we adopt one of the most commonly measured components of creativity from computational creative systems: novelty \cite{grace2015modeling}. Novelty is associated with measuring how different an artifact is compared to another set of artifacts \cite{grace2015modeling}. The novelty can be based on a comparison with a universal set of artifacts, which we will call a \textit{universal measure}, or on a set of artifacts that the user has previously experienced, which we will call a \textit{personal measure}.  In this paper, we use a universal measure based on a large dataset of labelled sketches and deep learning that enables two kinds of representation: one that enables a measure of visual similarity and one that enables a measure of conceptual similarity. From these metrics, we have constructed a universal composite measurement of novelty that is a combination of the distance between feature vectors in the visual space and the conceptual space. 

We hypothesize that, when a system provides stimulus in the form of design concept responses that are highly novel to the user's design, it leads to more transformative creative outcomes. In these cases, the designer is able to draw upon distant visual and semantic features to inspire their creative process, such as adding features from another design domain. In contrast, when the system displays stimulus design concepts that are less novel to the user's design, it corresponds to less creative outcomes. The features of similar designs do not provide highly novel input to the process, leading to design iterations that share many attributes with the designer's original sketch. To explore this hypothesis we performed a user study utilizing a Wizard of Oz system to see how altering the novelty of the AI agent's response affected the creativity of the user's response. Participants experienced three conditions: low, intermediate, and high novelty in the system's response. After the sketching experience, participants were interviewed and surveyed to determine how the AI agent's responses affected their creativity. We found that, based both on our quantitative and qualitative results, the high novelty conceptual shifts stimulated more creative thinking than the low novelty ones.

\section{Related Research}
Over the last few decades, digital tools have been introduced as a way to support design creativity \cite{johnson2009computational}. These tools offer a variety of functions that allow designers to share their digital sketches and suggest new ideas to facilitate creativity. More recently, intelligent systems have been developed that enable collaboration with designers in real time. These systems, also referred to as computational co-creative systems, work alongside human users to encourage their creativity, support inspiration, and stimulate the user to continue creating. ViewPoints AI \cite{jacob2013viewpoints} is an example of an artistic co-creative system that has applications in dance and theater. It uses a compositional technique that perceives and analyzes human movements and gestures to facilitate an AI response in real time. Morai Maker \cite{guzdial2019friend} is an example of a co-creative game level design tool that assists users in authoring game level content. 

Co-creative sketching systems are an active area of research in the computational creativity community. One such example is the Drawing Apprentice, which is a co-creative drawing partner that collaborates with users in real time \cite{davis2015drawing}. The system uses sketch recognition to identify objects drawn by the user and selects a complementary object to display on the screen. Complementarity is defined by the semantic distance between the user's sketched object and the target object. DuetDraw \cite{oh2018lead} is another example of a co-creative sketching tool that works alongside the user by recognizing what the user draws and drawing related content to complete a shared scene. In our work, we use visual and conceptual similarity to select an object from a distinct category to be drawn on the screen in order to support the design process. Instead of selecting a sketch from the same conceptual category, such as Drawing Apprentice, the CSP uses a computational model of conceptual shifts \cite{karimi2018creative} to determine an appropriate target sketch from a dataset. 

Conceptual shifts in design can occur when a sketch of one concept is recognized as being similar to a sketch of another concept \cite{karimi2018creative}. Identifying and capitalizing on conceptual shifts is an important component of the design process, as it allows designers to perceive their design ideas from  different perspectives. There are two modes of perception that have been defined in design: \textit{seeing-that} and \textit{seeing-as} \cite{suwa1997architects}. Seeing-that refers to the concrete properties of a sketch and their function in the overall design, whereas seeing-as refers to interpretation, in which sketch elements can be considered through multiple perspectives. Conceptual shifts have the potential to inspire designers to adopt the seeing-as mode of perception, exploring how their emerging design could be connected to a variety of distinct concepts presented as stimuli.

Identifying conceptual shifts could also help users overcome design fixation \cite{purcell1996design}. Designers often have a hard time disengaging from the ideas they developed and learned over time. This effect, called fixation, may be reduced by presenting designers with a sketch of another object that shares some visual and conceptual information. We presume that, when presenting a conceptual shift successfully triggers seeing-as perception, a designer could be distracted from fixation, and potentially develop novel contributions to their design. This could lead to the discovery of innovative solutions for a design task.

The study of creative design has lead to a characterization of different types of creativity. \citeauthor{gero2000computational} \shortcite{gero2000computational} has introduced six forms of design creativity that can form the basis for computational aids: combination, exploration, transformation, analogy, emergence, and first principles. Combination happens when two distinct design concepts are added. Exploration relates to changing some variable values associated with a design concept. Transformation involves altering one or more variables of a design concept through external processes. Analogy is characterized by mapping between structural elements of two dissimilar objects. Emergence occurs when extensional properties of a design concept are identified beyond the intentional ones. First principles use computational knowledge to relate function to behaviour and behaviour to structure. The CSP introduced in this paper can be considered a computational aid to design that can support the first four of these forms of creativity in a co-creative design context: combination, exploration, transformation, and analogy.

\section{Quantifying Conceptual Shifts}

Quantifying conceptual shifts is challenging because concepts are not typically represented or evaluated numerically. Our premise is that the larger the shift, the more creative the resulting design. In order to quantify the scale of a conceptual shift between two sketches (in our case the user's sketch and the system's proposed response), we need a representation space in which we can measure similarity or novelty. The more similar the second sketch is to the first, the less novel the second item is and (we hypothesize) the less likely that it will trigger a conceptual shift. When the two items are less similar, the more novel the stimulus and (again, we hypothesize) the more likely it will result in a conceptual shift.

We focus on novelty in generating conceptual shifts because it has been shown to be a key component in predicting creativity \cite{grace2015modeling}. The assumption in measuring novelty is the existence of a representation that allows objective measurement of difference. In \cite{grace2015modeling}, the corpus of designs in the design space were represented as a set of features that formed the basis for correlation and regression analysis. The feature set was extracted from a database in which the information about the designs was manually entered as a set of features with categorical and numerical values. This representation enabled various ways to measure novelty, but not a single novelty score. 

In the CSP, we measure novelty by comparing two sketches: an initial sketch presented by the user and a second sketch selected from a large dataset of sketches. Novelty is a combination of two components: the visual similarity based on the visual data and the conceptual similarity based on the label for the sketch. We use deep learning models to extract a vector representation in two design spaces: a visual space using a large dataset of sketches, and a semantic space using a word embedding model.  We consider the novelty to be a combination of the classification of visual novelty in the visual space and conceptual novelty in the word embedding space. 

We classify novelty into three categories: low, intermediate, and high. Low novelty occurs when two sketches share a large amount of visual and conceptual information, intermediate novelty is when two sketches share some visual and conceptual information, and high novelty occurs when two sketches share little visual and conceptual information. We presume that low novelty lies within the expectation of the user, and that the system's response might be most likely to help the designer add more details to their initial design. Intermediate novelty could instead inspire the designer to explore possible new design ideas associated with their initial design. High novelty has the potential to widen the user's thinking process, making it more likely to help them incorporate new design features from a completely different design space. Based on this presumption, we hypothesize that increasing the novelty of the CSP stimulus will correlate with more creative outputs.

\section{Conceptual Shift Algorithm}

In this section we describe an AI model of conceptual shifts. The model selects an object from a database of sketches to be displayed on the canvas as a stimulus during a co-creative session. Our model has two components: visual similarity and conceptual similarity. Visual similarity recognizes pairs of sketches from distinct categories that share some underlying visual information. Conceptual similarity identifies the semantic similarity between the labels of the sketches. 

Figure 2 shows the computational model the AI agent uses to select a sketch of the desired level of novelty in response to the user's input. The visual similarity module computes the distances between the cluster centroids of distinct categories and maps the user's input to the most similar sketches from categories to which it does not belong. The conceptual similarity module takes the pairs of selected category names from the previous step and computes their semantic similarity. In this section, we describe how CSP generates a numerical value for visual and conceptual similarity and determines the conceptual shift candidates based on high, intermediate, and low novelty.

\begin{figure*}[t]
    \centering
    \includegraphics[width=16.5cm,height=11.5cm]{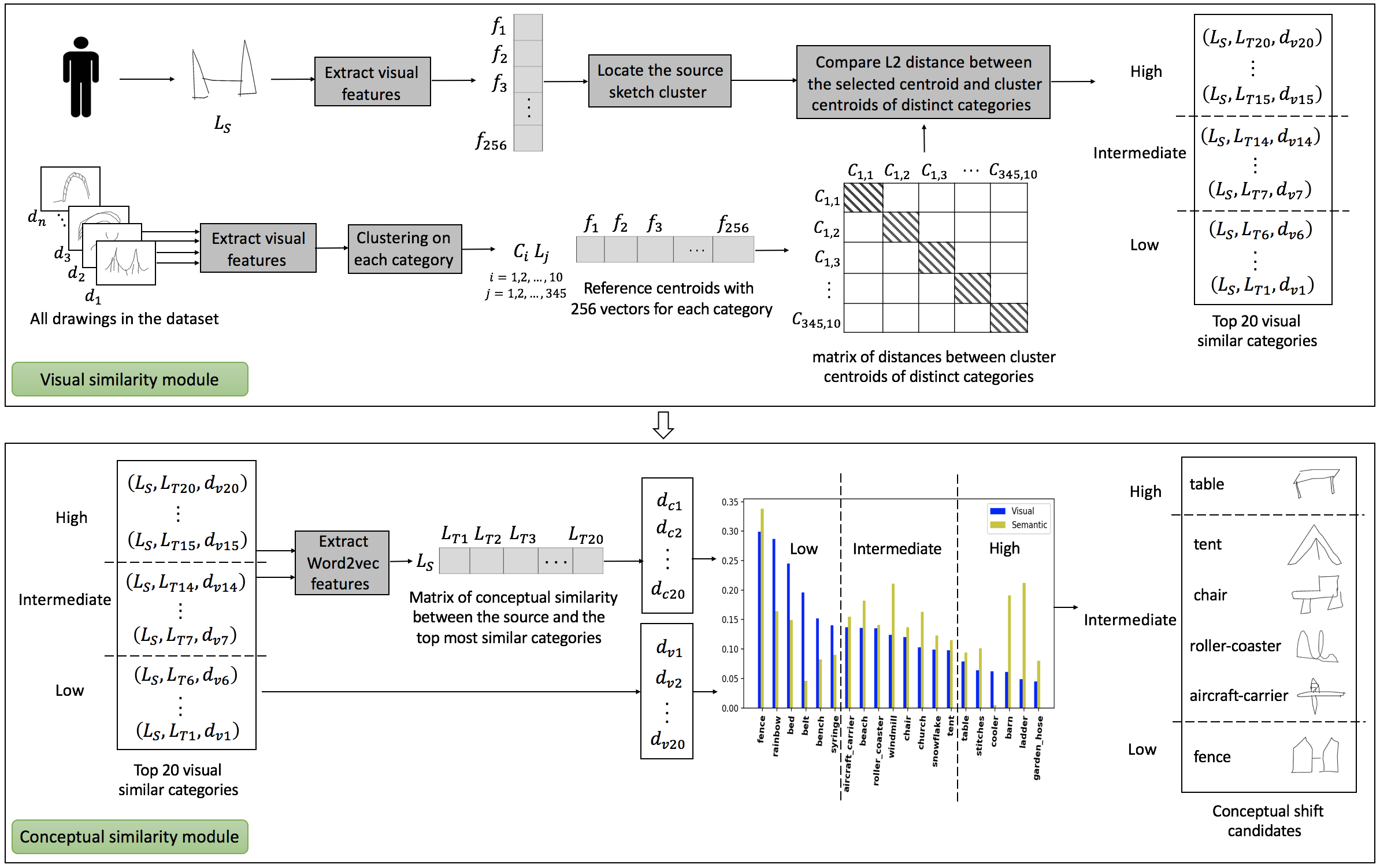}
    \caption{Computational steps for identifying conceptual shifts. Top: Identifying visually similar categories to the user's input. Bottom: Balancing visual similarity with conceptual similarity and identifying conceptual shifts with high, intermediate, and low novelty.}
    \label{fig:my_label}
\end{figure*}

\subsection{Visual Similarity Module}

The visual similarity module uses a large public dataset of human-drawn sketches, called QuickDraw! (QD) \cite{jongejan2016quick}, with more than 50 million labeled sketches grouped into 345 categories. In preparation for calculating visual similarity, we have 2 steps: a learning step and a clustering step. In the learning step, the sketches are used to build a vector representation of the sketch's features. In the clustering step, we use the resulting feature vectors for sketches in each category to create clusters of visually similar sketches. This process provides a feature vector representation for calculating the novelty between the user's initial sketch and sketches in the QD dataset using visual similarity.

\subsection{Deep Learning Model of Sketches for Visual Similarity}

As in the case of natural images, sketches can also be processed as a grid of pixels, $(h,w,d)$, in which $h$ is the height, $w$ is the width, and $d$ is the number of channels. However, in this case, $d$ will be 1 because the sketches are monochrome. To develop a representation for visual similarity we employed a convolutional neural network (CNN) model due to their success in providing high level visual information and discriminating visual appearances, such as shapes and orientations \cite{lecun2015deep}. We started with a pre-trained model, VGG16 \cite{simonyan2014very}, with 13 convolutional layers, two fully connected layers, and a softmax output layer. The model is primarily trained on the ImageNet dataset \cite{deng2009imagenet} that contains more than 20 million labeled natural images. We then fine-tune this model on the QD dataset with the objective of classifying a sketch into one of the 345 categories.  We use 30,000 training samples and 10,000 validation samples per category, and trained for 1.5 million training steps. Observation shows that the accuracy reaches 52.1\% after 1 million steps and remains the same afterwards. We extract a neural representation of each sketch by taking the output of the first fully connected layer, for 4096 values per sketch. However, this model has low accuracy and a high computational cost because of the large number of parameters in the VGG16 architecture and processing sketches as a grid of pixels.

In order to solve this problem, we tried another representation of sketches: a sequence of pen strokes, inspired by the work done by Ha and Eck on Recurrent Neural Network drawing \cite{ha2017neural}. In this case, each stroke is a list of points with 3 elements: ($\Delta$x, $\Delta$y, $p$). $\Delta$x and $\Delta$y are the coordinates with respect to the previous point, and $p$ is a binary number that determines whether the stroke is drawn or not (i.e. just moves the pen). Here we use a deep learning model called Convolutional Neural Network-Long Short Term Memory (CNN-LSTM) \cite{RNNquickdraw}. The model has three one-dimensional convolutional layers and three LSTM layers. We train the model from scratch on the QD dataset with the same objective, training, and validation samples as the CNN-only model. Results show that, after 1 million training steps, accuracy reaches 73.4\% and remains the same afterwards. Each sketch is represented by the last LSTM layer, for 256 values per sketch. Table 1 summarizes the results for accuracy and the average-per category inference time for both models. Accuracy measures a true positive rate, while inference time represents the total amount of time it takes to extract features from all sketches of a category.  The CNN-LSTM model is clearly both faster and more accurate, and we use it hereafter.

\subsection{Clustering visually similar sketches in each category} The sketches in a category exhibit a large variability visually. For our visual similarity measure to be meaningful, we group the sketches in each category into clusters and use the feature vector of the cluster centroid as the representative sketch. This process is a form of denoising, where the intra-cluster variability is suppressed. We perform clustering using a K-means algorithm and determine the optimal number of clusters via the elbow method. By analyzing the variance versus the number of clusters, we observed that for most categories the optimal number of clusters is between 7 and 12---we set the number of clusters to 10 across all categories. The distances between the cluster centroids from distinct categories are computed and stored in a matrix of size $3450 \times 3450$: 10 clusters of sketches for each of 345 categories.

Given the source sketch and label from the user, $L_S$, we first extract visual features using the pre-trained CNN-LSTM model that produces 256 values. We then locate the representative cluster within its category (according to the label of the user's sketch) by selecting the closest centroid based on the L2 (i.e. Euclidean) distance. Using the distance matrix, we then select the top 20 most visually similar target clusters from other categories, $L_T$, as the ones with minimum distance from the representative cluster. The similarity is computed as $1- d_v$, where $d_v$ is the Euclidean distance normalized across the most visually similar candidates. As the similarity values for the selected target sketches change smoothly, we classify those that fall in the top 33rd percentile of the distribution as low novelty, between 33rd and 66th percentile as intermediate novelty, and above 66th percentile as high novelty.

\subsection{Conceptual Similarity Module}

The conceptual similarity module uses a word embedding model \cite{mikolov2016word2vec} trained on the Google News corpus with 3 million distinct words. The visual similarity module provides a set of candidate sketches to the conceptual similarity module based on the categories of low, intermediate, and high novelty. We extract the word2vec word embedding features \cite{mikolov2016word2vec} from these category names. The similarity between the category of the source sketch and the selected target sketch is computed as $1- d_c$, where $d_c$ is the cosine distance between the feature vectors of category names. The larger number indicates that the two sketch categories are more likely to appear in the same context, whereas a smaller number indicates that the two are less associated with each other. In order to determine the conceptual shift categories, we select those where the visual and conceptual similarity are both high, medium, or low. This is done by selecting candidates for which the difference between visual and conceptual similarity values are below 0.05 and the overall similarity component is computed as the average of visual and conceptual values. 

\begin{table}[t!]
\centering
\def\arraystretch{1.5}
\begin{tabular}{|p{0.9in}|p{0.9in}|p{0.9in}|} \cline{2-3}
\multicolumn{1}{c|}{} &\hfil \textbf{VGG-16} & \hfil \textbf{CNN-LSTM} \\ \hline
 \hfil Accuracy &  \hfil 52.1\% &  \hfil 73.4\% \\ \hline
\hfil Inference time &  \hfil 18,000S & \hfil 960S \\ \hline
\end{tabular}
\caption{Classification accuracy and the inference time using two different deep learning models.}
\end{table}

\section{User Study}

We conducted a user study to evaluate the effectiveness of our conceptual shift model in a co-creative design session. We investigated how the novelty of the system's response could inspire user creativity and correspond to different types of design behaviors. Our hypothesis is that increasing the novelty of the system's response can help designers add new features and/or functions from another design space to their initial drawing, thus leading to more creative outcomes. By contrast, when the system is in the low novelty condition, the designer is presented with the similar features to the initial drawing, which leads to less creative outcomes.  

In this study, we used a within-subjects design, such that each participant experienced three conditions with a two-minute break between them. In the first condition the design task is a chair, and the system produces a result that is highly novel with respect to the participant's sketch. In the second condition the design task is a streetlight, and the system produces a result associated with intermediate novelty. In the third condition the design task is a bridge, and the system produces a result that is classified as low novelty. Participants were not aware whether they were in a high, intermediate, or low novelty condition. A context is provided to help guide each design task, such as ``draw a streetlight for safety at night on a city street of a small town.'' When the system's output object is presented to the user, it is accompanied by a label indicating what the object is. Each design task takes approximately 7 minutes. The order of the three conditions for each participant was randomized to account for any ordering effects.

We used an online sketching tool, called SketchTogether \cite{SketchTogether}, that enables multiple users to contribute to a shared canvas in real time. This application allowed us to run a Wizard-of-Oz interaction for the user study in which we used the results of the deep learning model for determining high, intermediate, and low novelty sketches, but a person performed the interaction of placing the selected sketch on the shared canvas. Participants underwent a 5-minute training session that included an explanation about the interface tool and the design tasks. After training, participants are asked to start the first design task. The instruction given to the participants were to draw an object according to the design task and iterate on that drawing based on inspiration from the system's response to their sketch. Following each experimental condition, we asked participants Likert scale survey questions associated with that design session. The questions we asked after each task were:

\begin{figure*}
    \centering
    \includegraphics[width=17.5cm]{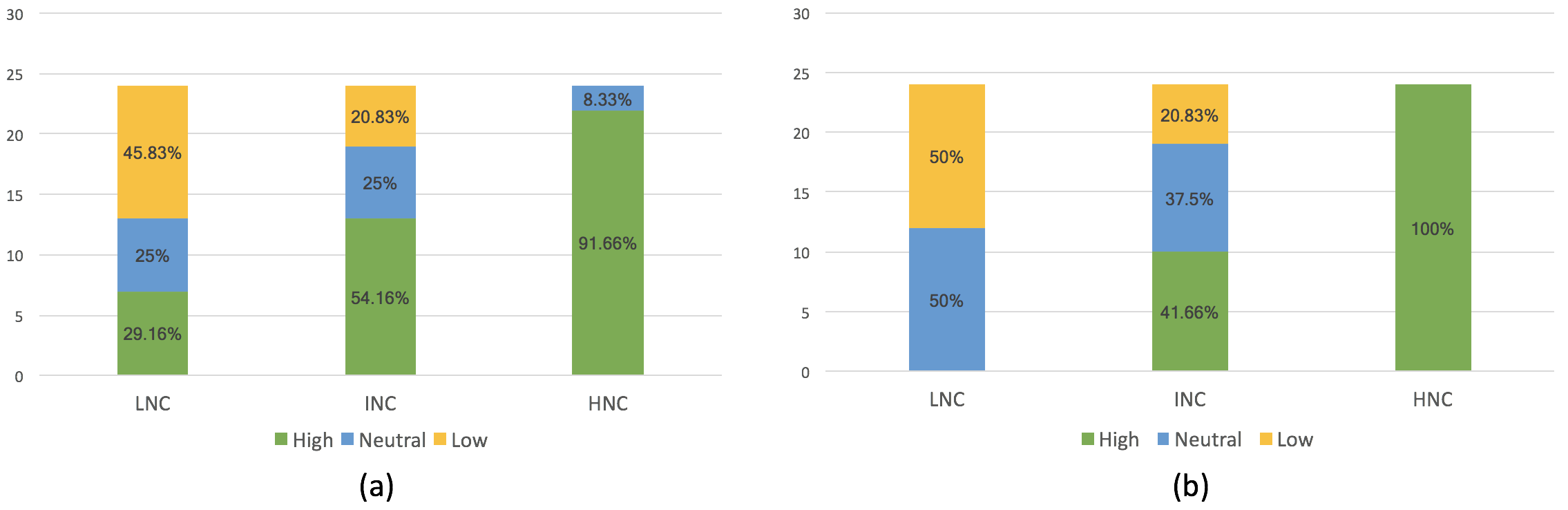}
    \caption{The total percentage of high, intermediate, and low survey responses for (a) inspired creative ideas, and (b) led to different design.}
    \label{fig:my_label}
\end{figure*}

\begin{enumerate}
\item Did the system's sketch response inspire you to come up with creative ideas for your design objects?
\item Did the system's sketch response lead you to come up with a different type of design object?
\end{enumerate}

The answers to the survey questions were recorded for quantitative analysis. After the last design session, we asked participants the following questions in an interview:

\begin{enumerate}
\item How did the sketches presented by the system affect your creative process?
\item Was it more helpful when the sketches presented by the system were more or less similar to your input?
\item In which of the three design tasks did the system's sketch inspire you most?
\item Do you have any comments for participating in this study?
\end{enumerate}

The answers to the interview questions were used for qualitative analysis. The entire session for each participant took almost 30 minutes. 

\section{Results}

The user study included 24 participants recruited from the College of Architecture at a public university in North America. Gender distribution was 15 males and 9 females. The criterion for participating was whether students perform sketching frequently for their design practice. We recorded survey and interview responses for all participants. In this section, we describe our analysis based on the participants' responses in order to investigate our hypothesis. 

\section{Quantitative Analysis} We compared the results from the user's feedback on the three design tasks associated with high, intermediate, and low novelty conditions. We grouped the responses into high, neutral, and low ratings: 4 and 5 are considered high, 3 is neutral, and 1 and 2 are low. For each condition we count the number of ratings based on this grouping.

\subsection{Analysis of creative ideas}

Participants were asked to rate the responses provided by the system after each design session. With this question, we aimed to understand whether increasing the novelty of the system's response inspired their creative thoughts. We found that 91.66\% of the participants thought that the system's response inspired creativity when the system was in the high novelty condition (HNC) compared to 29.16\% in the low novelty condition (LNC). These results indicate that when the system's response is more novel with respect to the user's sketch (HNC), it is associated with more creative outcomes, which may encourage the user to come up with new design ideas for their initial drawing. When the system was in intermediate novelty condition (INC), 54.16\% of the participants were highly inspired by the system's response. Figure 3a shows the distribution of the ratings for the three conditions.

\subsection{Analysis of design object inspiration}

Transformational creativity happens when a designer changes one or more structural variables of the current design object to produce new variables \cite{gero2000computational}. This implies that the system's response has the potential to inspire the user to transform some features of a design concept by adding new features from another design space related to the system's response. We explored whether increasing the novelty of the system's response can lead to transformational creativity in which the participant's designed object significantly deviates from their initial sketch. All participants rated high in response to changing their design when the system was in HNC. This indicates that when the system's response was less similar to the participant's input (HNC), they were able to transform their initial sketch. By contrast, when the system was in LNC, none of the participants reported that the system helped them come up with a different type of design object. when the system was in INC, 41.66\% of the participants rated high in response to changing their design and 58.33\% rated low or neutral (see Figure 3b).

\section{Qualitative Analysis}

To understand how the novelty of the system's response can help designers come up with creative ideas for their initial task we analyzed the participants responses to the interview questions conducted after the design tasks were complete. We aimed to explore the relationship between stimulus novelty and design thinking.

\subsection{Thematic Analysis} 

We performed a thematic analysis of the responses the participants gave to the interview questions. Overall, three main themes were found from the interview answers. 

\begin{itemize}
\item The tool helps with the design process
\item High novelty helps changing the design  
\item Low novelty helps completing the design
\end{itemize}

In the following section, we elaborate on each of these themes.

\subsection{Supporting the design process}

Most participants found the tool useful, as it can help with the design-thinking process as well as iterating and generating new design ideas. P11 exemplifies how the sketching tool helped their design process, ``\textit{The sketches presented after I did my initial sketch, change the creative process, making me think of different object and using that design philosophy and then the second object to affect the first.''} This participant described how the system's output sketch helped them think of different design ideas and iterate on their initial design sketch. This demonstrates that the tool generally supports the iterative nature of the early design process. Additionally, P14 comments: ``\textit{it sort of help[ed] me to see how I think about design, like they teach us just to design, I never really thought about how I go about that process of designing and so having this sort of precedent to work with is more useful to me.''} This participant shows the role such a tool could play in design education. It helps to provide precedents that can inform the design process and inspire additional thinking on the topic.

P4 described how helpful the system is when they say, \textit{``I think the system's response is very helpful, because it gives me a leverage on adding to my initial design or just give me some clue or hint to change my design to make it better.''} Here, the participant comments about how the tool helps them iterate on their design by adding or changing different elements of the initial sketch based on the `clues' or `hints' provided by the system's output. P5 agrees with this sentiment when they said, \textit{``the way that we communicate is great because you add something and I am going to redesign it and so it's great.''} This participant focused on the communication channel established between the user and system, and described how this channel helped in the redesign process. In a similar vein, P25 describes how \textit{``it kind of guided me through some conventional ways of improving my design''} which shows how the tool serves to shepherd users through the design process by providing new avenues to explore and inspiration to change the user's initial design.

\subsection{High novelty inspires changing the design}

We found that high novelty conceptual shifts inspire participants to change the overall shape of their design by adding new features from another design space related to the target sketch. In this condition, 21/24 reported that it is more inspiring when the system's response is less similar to their initial design. P11 commented: \textit{``I think to create an interesting result it was more helpful to have a dissimilar object as opposed to a similar, because it allows you to change the form and different ideas instead of just kind of a similar shape affecting it.''} This participant indicates that when the system's response is less similar to their initial design (high novelty condition), it helps to change the structure, such that it is possible to incorporate different ideas from the target sketch. Similarly, P10 commented: \textit{``It was easier to make changes when it was more different. I think when something is already similar sometimes my brain already has a same set of ideas, but when I am presented with something different the contrast helps me to generate a new idea.''} This participant was able to come with a new idea when he/she was presented with a sketch that was less similar to the initial drawing. 

When P16 was presented with a sketch of an aircraft-carrier after designing a chair, they described how the system's sketch opened up new possibilities for them, \textit{``The aircraft-carrier may have chairs but it doesn't elicit specific form especially giving the prompt that is going to be at the kitchen table. Thinking about new possibilities that can happen definitely opens the new design criteria.''} This example shows that the chairs of the aircraft-carrier introduced new design criteria that inspired the participant to sketch a new kitchen chair with the features of aircraft-carrier seats, such as more comfort. Additionally, when P21 was presented with a sketch of a speedboat after designing a chair, they also found new possibilities in the design space, \textit{``The relationship between the two, even though they are used both in the same task or same function because of the difference that one is on water, one needs to be outdoor, the different needs and purposes between the two was influencing me better to create something new between them.''} Similarly, P22 used the features of the system's response to reason about their initial sketch, \textit{``The aircraft, because of its curves and the materiality, so thinking about the skin of the material, maybe thinking about its curves so that led me to think about the curves which maybe helped me to think of armrest.''} In this example both the structure and the concept of the target sketch inspired the participant to change the shape to be curvy as well as adding new functionalities such as armrests. 

\subsection{Low novelty helps complete the design}

Overall, 3/24 participants commented that it is more helpful when the sketch that is presented to them is more similar to their initial drawing (low novelty condition). P4 explains why the sketch of fence that was highly similar to their initial drawing of bridge was more helpful, \textit{``because there were clear features and structures that could help by adding, mainly the similar features.''} In this case, the participant preferred to finalize the original drawing by adding more details and structures rather than changing the existing features. Similarly, P9 commented: \textit{``I like the product of end results when stuff [is] more similar. Because I could pull from the profile of fence and add to the bridge...So, you take something from it and add it to your design.''} From both P4 and P9, we can conclude that when the system is in low novelty mode the designer mainly adds more details to the initial drawing rather than transforming the shape or adding new features to the drawing. Most participants found the low novelty condition less helpful. For instance, P12 described how they liked less similar designs, \textit{``I would say it was more helpful when it was less similar because then you are not just copying the instances from the other design.''} P8 agreed with this sentiment when they said: \textit{``high similarity is kind of within my expectation.''} 

In both cases of P8 and P12, the low novelty conceptual shift designs do not help to significantly change the original drawing. Instead, they are used to combine some elements of the two sketches. P13 echoes this general viewpoint when they said: \textit{``I think if you are presenting something that is almost exactly the same, you are going to introduce the same idea again.''} Similar to P8, this participant also emphasizes that low novelty conceptual shifts are within their expectation. P22 also commented: \textit{``I feel that similar designs didn't give me as much creative freedom.''} These examples demonstrate that low novelty conceptual shifts may help to combine the elements of the two sketches, rather than encouraging the user's creative thoughts. Both likely have a role in co-creative design systems, serving different purposes.

\section{Conclusion}

This paper presents a computational model of conceptual shifts for a co-creative design system called the Creative Sketching Partner. The tool is meant to inspire design creativity by presenting a sketch of a distinct category that shares some visual and conceptual information with the user's input sketch. We describe the role of deep learning in creating a representation space for measuring distance between the visual and conceptual features of a sketch. We have detailed the process for classifying potential response sketches as low, intermediate, or high novelty with respect to the designer's sketch. A user study is presented in which the participants are given a design task and then experience three different versions of the tool: low, intermediate, and high novelty responses. Both quantitative and qualitative results from the user study demonstrate that the high novelty conceptual shift designs inspire creative thinking more than the low novelty condition. 

\section{Acknowledgements}

The research reported in this article is funded by NSF IIS1618810 CompCog: RI: Small: Pique: A cognitive model of curiosity for personalizing sequences of learning resources.

\bibliographystyle{iccc}
\bibliography{iccc}

\end{document}